# A Gravitational Tractor for Towing Asteroids

Edward T. Lu and Stanley G. Love

*NASA Johnson Space Center*

**We present a concept for a spacecraft that can controllably alter the trajectory of an Earth threatening asteroid using gravity as a towline. The spacecraft hovers near the asteroid with thrusters angled outward so the exhaust does not impinge on the surface. This deflection method is insensitive to the structure, surface properties, and rotation state of the asteroid.**

The collision of an asteroid as small as ~200 m with the Earth could cause widespread damage and loss of life[1]. One way to deflect a threatening asteroid is to dock a spacecraft to the surface and push on it directly[2]. The total impulse needed for rendezvous and deflection is too large for chemical rockets, but is achievable by 20 ton class nuclear-electric propelled spacecraft proposed by NASA[2]. Regardless of the propulsion scheme, a docked asteroid tug needs an attachment mechanism since the surface gravity is too weak to hold it in place. Asteroids are likely to be rough and unconsolidated, making stable attachment difficult. Furthermore, most asteroids rotate, so an engine anchored to the surface would thrust in a constantly changing direction. Stopping the asteroid's rotation, reorienting its spin axis[3], or firing the engine only when it rotates through a certain direction adds complexity and wastes time and propellant.

Our suggested alternative is to have the spacecraft simply hover above the surface. The spacecraft will tow the asteroid with no physical attachment using gravity as a towline.

The thrusters must be canted outboard to keep them from blasting the surface (which would reduce net towing force and stir up unwanted dust and ions). This scheme is insensitive to the asteroid's poorly known surface properties, internal structure, and rotation state. The spacecraft need only stationkeep in the direction of towing while the asteroid rotates beneath it. The engines must be actively throttled to control vertical position since the equilibrium hover point is unstable. Horizontal position will be controlled by differential throttling of engines on opposite sides of the spacecraft. The spacecraft can be made stable in attitude by designing it like a pendulum, with the heaviest components hanging closest to the asteroid and the engines further away.

The thrust required to balance the gravitational attraction is given by

$$T\cos[\sin^{-1}(r/d)+\phi] = GMm/d^2 = 1.12(\rho/2g/cm^3)(r/d)^3(m/2\times10^4 Kg)(d/100m).$$

Thus a notional 20 ton spacecraft with $\phi = 20°$ hovering one half radius above the surface ($d/r = 1.5$) can tow a 200m diameter asteroid ($r = 100m$) with density $\rho = 2g/cm^3$ provided it can maintain a total thrust $T = 1$ N. The velocity change imparted to the asteroid per year of hovering is $\Delta v = 4.2\times10^{-3}(m/2\times10^4 Kg)(d/100m)^{-2}(m/s)(yr)^{-1}$. Because $\Delta v$ is largely independent of the asteroid's detailed structure and composition, the effect on the asteroid's orbit is predictable and controllable, as needed for a practical deflection scheme.

The mean change in velocity required to deflect an asteroid from an Earth impact trajectory is $\sim 3.5\times10^{-2}/t$ m/s where $t$ is the lead time in years[4]. Thus, in the example above, a 20 ton gravitational tractor can deflect a typical 200m asteroid, given a lead time of about 20 years. The thrust and total fuel requirements of this mission are well within the capability of proposed 100kW nuclear-electric propulsion systems[2], using about 4 tons of fuel to accomplish the typical 15 km/sec rendezvous and about 400 Kg for the actual deflection. For a given spacecraft mass, the fuel required for the deflection scales linearly with the asteroid mass. Deflecting a larger asteroid requires a heavier spacecraft, longer time spent hovering, or more lead time. However, in the special case where an asteroid has a close Earth approach followed by a later return and impact, the change in velocity needed to prevent an impact can be many orders of magnitude smaller if applied before the close approach[5]. For example, the asteroid 99942 Apophis (2004 MN4), a 320m asteroid that will swing by the Earth at a distance of ~30000km in 2029, has a small $10^{-4}$ probability of returning to strike the Earth in 2035 or 2036[6]. If it indeed is on a return impact trajectory, a deflection $\Delta v$ of only ~$10^{-6}$ m/s a few years before the close approach in 2029 would prevent a later impact (Carusi, personal communication). In this case, a 1 ton gravitational tractor with conventional chemical thrusters could accomplish this deflection mission since only about 0.1 Newtons of thrust are required for a duration of about a month. Should such a deflection mission prove necessary, a gravitational tractor spacecraft offers a viable method of controllably steering asteroid 99942 Apophis away from an Earth impact.

1. Chapman, C.R. The Hazard of Near-Earth Asteroid Impacts on Earth. *Earth and Planetary Science Letters* **222,** 1-15 (2004).

2. Schweickart, R.L., Lu, E.T., Hut, P., and Chapman, C.R. The Asteroid Tugboat. *Scientific American,* 54-61 (2003).

3. Scheeres, D.J., and Schweickart, R.L. The Mechanics of Moving Asteroids. *AIAA paper 2004-1446.* (2004).

4. Chesley, S.R., and Spahr, T.B. Earth Impactors: Orbital Characteristics and Warning Times, *Mitigation of Hazardous Comets and Asteroids, editors Belton, M.J.S. et al.* Cambridge University Press, 22-37 (2004).

5. Carusi, A., Valsecchi, G.B., D'Abramo, G., and Boatini, A. Deflecting Near Earth Objects (NEO) In Route of Collisions with the Earth. *Icarus* **159,** 417-422 (2002).

6. JPL Sentry Impact Risk Page, *http://neo.jpl.nasa.gov/risk/*

Figure 1 Caption: Gravitational tractor towing geometry. The (assumed spherical) asteroid has radius $r$, density $\rho$, and mass $M = \frac{4}{3}\pi r^3 \rho$. The spacecraft has mass $m$, total thrust $T$, and exhaust plume half width $\phi$. It hovers at distance $d$ from the asteroid's center, its net thrust balancing its weight. The thrusters are tilted outward by $\sin^{-1}(r/d) + \phi$ to avoid exhaust impingement.

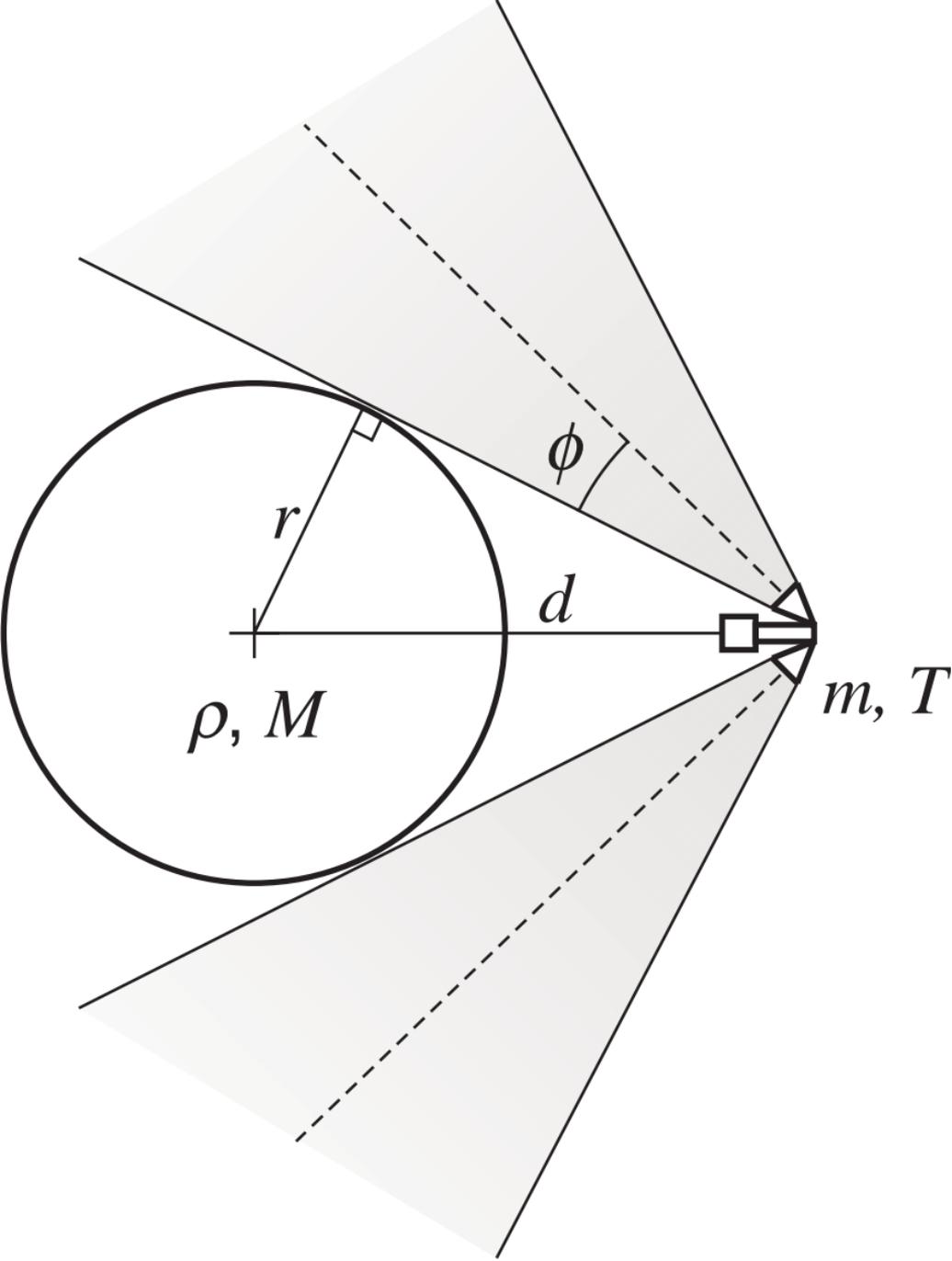